\documentclass[twocolumn,amsmath,amssymb,aps,prb,floatfix,nofootinbib,superscriptaddress]{revtex4}
\usepackage{graphicx,graphics,times,color}
\usepackage{bm}
\usepackage{longtable}
\usepackage{color}
\usepackage{lipsum}
\usepackage{epstopdf}
\usepackage{amsmath}

\def\be{\begin{equation}}
\def\ee{\end{equation}}
\def\ba{\begin{eqnarray}}
\def\ea{\end{eqnarray}}
\def\la{\langle}
\def\ra{\rangle}

\def\h{\hskip 1cm}

\begin{document}

\title{Universal single-frequency oscillations in a quantum impurity system after a local quench}

\author{Abolfazl Bayat}
\affiliation{Department of Physics and Astronomy, University College
London, Gower St., London WC1E 6BT, United Kingdom}

\author{Sougato Bose}
\affiliation{Department of Physics and Astronomy, University College
London, Gower St., London WC1E 6BT, United Kingdom}

\author{Henrik Johannesson}
\affiliation{Department of Physics, University of Gothenburg, SE 412 96 Gothenburg, Sweden}

\author{Pasquale Sodano}
\affiliation{International Institute of Physics,
Universidade Federal do Rio Grande do Norte,
59078-400 Natal-RN, Brazil,
and \\
Departamento de F�isica Te\'{o}rica e Experimental, �
Universidade Federal do Rio Grande do Norte,
59072-970 Natal-RN, Brazil}
\affiliation{INFN, Sezione di Perugia, Via A. Pascoli, 06123, Perugia, Italy}

\begin{abstract}
Long-lived single-frequency oscillations in the local non-equilibrium dynamics of a quantum many-body  system is an exceptional phenomenon. In fact, till now, it has never been observed, nor predicted, for the physically relevant case where a system is prepared to be quenched from one quantum phase to another. Here we show how the quench dynamics of the entanglement spectrum may reveal the emergence of such oscillations in a correlated quantum system with Kondo impurities. The oscillations we find are characterized by a single frequency.
This frequency is independent of the amount of energy released by the local quench, and scales with the inverse system size.
Importantly, the quench-independent frequency manifests itself also in local observables, such as the spin-spin correlation function of the impurities.
\end{abstract}


\maketitle

\section{Introduction}
The dynamics of an isolated quantum many-body system following an instantaneous change of a Hamiltonian parameter {\em (quantum quench)} is a topic of growing  interest \cite{Eisert-Ther-2014}. The problem touches on a multitude of subjects, from the foundation of quantum statistical physics to the engineering of  quantum states and devices \cite{Polkovnikov-NED-2011,Eisert2015}. A quench injects energy which disperses among the interacting degrees of freedom, and as time evolves local observables relax to their  equilibrium values \cite{Proukakis-Ther-2013}.  When the quench is {\em local}, with a sudden change of a local parameter in a Hamiltonian, the energy injected to the system is nonextensive. In such a case one may expect, and indeed finds \cite{Eisler-Ent-2007}, intermediate times at which wave propagation and reflection from boundaries can create slowly decaying oscillatory behaviour. This raises the question whether one could find a physically relevant model where, by tuning a pertinent parameter, the equilibration after a local quench of the Hamiltonian is strongly suppressed, or even totally eliminated.


Here, in point of fact, we show that, for the spin emulator \cite{Bayat-TIKM-2012} of the two-impurity Kondo model \cite{Jayprakash,ALJ}, a local quench into the Kondo-screened phase \cite{AffleckReview} induces the onset of long-lived oscillations. By analyzing the quench dynamics of the lowest eigenvalues of the entanglement spectrum \cite{Haldane-ES-2008,CL,Sanpera-SG-2012}, we find that the frequency $f_u$ of these oscillations is sharply determined if one tunes the impurity-spin interaction so that {\em all} spins become entangled with an impurity.  Remarkably, $f_u$ is independent of the local quench energy and scales as $f_u N$ remaining constant when the system size $N$ increases. Moreover, we find that the  frequency $f_u$ leaves its fingerprints on local observables, including the spin-spin impurity correlation functions.  As such, we believe that our finding is potentially relevant to future designs of spin-based quantum devices as the impurity spins do not seem to equilibrate despite being strongly connected to reservoirs.


\section{Model}
We consider  a spin chain emulator of the two-impurity Kondo model \cite{Bayat-TIKM-2012}  with two localized spin-1/2, each coupled to a frustrated spin-1/2 Heisenberg chain, and to each other via a Ruderman-Kittel-Kasuya-Yosida (RKKY) interaction. The Hamiltonian can be written as

\begin{figure} \centering
    \includegraphics[width=8cm,height=5cm,angle=0]{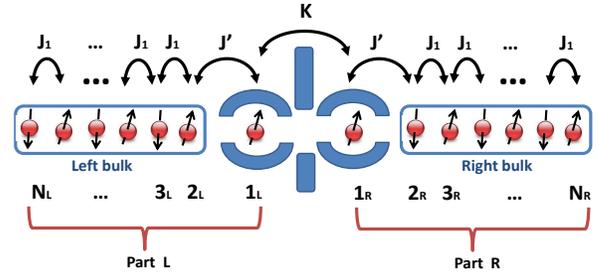}
    \caption{Schematic picture of the spin emulator of the two-impurity Kondo model.  The impurity coupling $J'$, together with the RKKY coupling $K$, determines the extension of the entanglement length $\xi(J',K)$. The strength of the RKKY coupling $K$ can be externally controlled; for instance, in a quantum dot,  by tuning gate voltages as indicated in the figure. For any $K\leq K_c$, in the Kondo regime, we define $J'_{opt}$ as the impurity coupling for which the entanglement length extends over the entire chain, i.e. $\xi(J'_{opt},K)\simeq N$. By quenching the RKKY coupling $K$ from $K_1$ in the RKKY regime to $K_2$ in the Kondo phase one induces the non-equilibrium dynamics being studied here. For the sake of simplicity the next-nearest coupling $J_2$ is not shown in the figure since we fixed $J_2/J_1=0.2412$.}
         \label{fig1}
\end{figure}

\begin{equation}  \label{Ham}
H=\sum_{m=L,R} \{ H_{imp}^m + H_{bulk}^m \} +H_{I},
\end{equation}
 where
\begin{eqnarray} \label{Hamiltonian}
H_{imp}^m & = & J^{\prime} \left( J_1 \boldsymbol{\sigma}_1^m \!\cdot \!\boldsymbol{\sigma}_{2}^m + J_2 \boldsymbol{\sigma}_1^m \!\cdot \!\boldsymbol{\sigma}_{2}^m \right) \cr
H_{bulk}^m &=& J_1 \sum_{i=2}^{N_m-1} \boldsymbol{\sigma}_i^m \!\cdot \!\boldsymbol{\sigma}_{i+1}^m  + J_2 \sum_{i=2}^{N_m-2} \boldsymbol{\sigma}_i^m \!\cdot \!\boldsymbol{\sigma}_{i+2}^m \cr
H_{I} &\!=\! &K J_1\boldsymbol{\sigma}_1^L \!\cdot \!\boldsymbol{\sigma}_1^R.
\end{eqnarray}
Here $m=L,R$ labels the left and right chains, with $\boldsymbol{\sigma}_i^m$ the vector of Pauli matrices at site $i$ on chain $m$.
The couplings $J_1$ and $J_2$ are nearest and next nearest neighbour couplings, respectively. The dimensionless parameter $J^{\prime}>0$ plays the role of an antiferromagnetic Kondo coupling between the impurities and their corresponding bulks and the dimensionless coupling $K>0$ measures the RKKY impurity interaction. The mapping to the Kondo model is valid for $J_2\leq  J_2^c$ (with $J_2^c=0.2412 J_1$) \cite{Sorenson-QIE-2007,Affleck-KondoSpinChain-2008} since in this interval the bulk excitations are massless (as in the full electronic version of the Kondo problem). As we will see, our analysis of the quench dynamics applies to the entire Kondo regime, although, for $J_2 < J_2^c$ a marginal coupling (in the sense of the renormalization group) produces logarithmic corrections which pollutes the numerical data \cite{Affleck-KondoSpinChain-2008,Deschner-2011}. In order to avoid this, we tune $J_2=J_2^c$ for which both $H_L$ and $H_R$ faithfully represent the spin sector of a single-impurity Kondo model.
Note that the terms $H_{imp}^m + H_{bulk}^m$ in Eq. (\ref{Ham}) each define a spin emulator of the single channel Kondo model, with labels $m=L,R$ \cite{Sorenson-QIE-2007}. A schematic picture of the model is given in Fig.~\ref{fig1}.

The ground state shows a quantum phase transition at a critical value $K=K_c$ of the RKKY coupling. For a small coupling $K$ the Kondo interaction dominates and each impurity spin gets screened by its bulk (Kondo phase), while in the opposite limit the impurities form a local singlet (RKKY phase) and decouple from the rest of the system \cite{ALJ}.

\section{Results}

\subsection{Entanglement spectrum}
Having partitioned the system into two parts, $L$ and $R$  (see Fig.~\ref{fig1}), one can write an arbitrary pure state in the orthogonal Schmidt basis \cite{Nielsen-Chuang-2000} as
\begin{equation}\label{Schmidt-decomp}
    |\psi_{LR}\ra=\sum_{n} \sqrt{\lambda_n} |L_n\ra \otimes |R_n\ra,  \ \ \lambda_n \geq 0,
\end{equation}
where the ordered set of real numbers $\lambda_1\geq \lambda_2 \geq ...$  form the entanglement spectrum. The Schmidt bases $\{ |L_n\ra\}_{n=1}^{2^{N_L}}$ and $\{
|R_n\ra\}_{n=1}^{2^{N_R}}$ diagonalize the reduced density matrices $\rho_L$ and $\rho_{R}$ of the left and right parts respectively. It has been shown that the two largest Schmidt numbers dominate the entanglement spectrum, with their difference, $\lambda_1-\lambda_2$, behaving as an order parameter at the quantum phase transition \cite{Sanpera-SG-2012,Bayat-SG-2014}. Notably, all levels of the entanglement spectrum contribute essentially to the von Neumann entropy
\begin{equation}\label{von-Neumann}
    s(\rho_L)=s(\rho_R)= -\sum_n \lambda_n \log(\lambda_n).
\end{equation}

In the Kondo regime $K\leq K_c$, the system supports a  length scale $\xi(J',K)$ which diverges at the critical point in the thermodynamic limit. It may be determined numerically by exploiting its interpretation as the length scale over which the two impurities are entangled with two identical blocks of spins on both sides  \cite{Bayat-TIKM-2012}. One thus finds, for a large but finite system with $N=N_L+N_R$ \cite{Bayat-SG-2014},
\begin{equation}\label{Kondo_length_xi}
    \xi(J',K)  \sim \frac{e^{-\alpha/J'}}{|K-K_c|^\nu + \mathcal{O}(1/N)} .
\end{equation}
Here $\alpha$ is a constant, with $\nu$ a critical exponent  taking the value $\nu \!=\! 2$ in the neighborhood of the critical point $K_c \sim e^{-\alpha/J'}$ \cite{Bayat-TIKM-2012}.
According to Eq.~(\ref{Kondo_length_xi}), for any given $K\leq K_c$  in the Kondo regime, it is always possible to find an optimal impurity coupling $J'=J'_{opt}$  such that $\xi(J'_{opt},K)=N$,  making each impurity entangled with {\em all} spins in its bulk. For $J' \ll J'_{opt}$ the length $\xi$ exceeds the length of the chain and Kondo screening does not take place \cite{AffleckReview}.

The time evolution of entanglement spectra following a quantum quench have been the subject of several recent investigations \cite{Cardy-Quench-2014,Zamora-Splitting-2014,DeChiara-EntSpec-2014,Vodola-EntSpect-2013,Calabrese-LQ-2007}. As we shall see next, the entanglement  dynamics of the present problem exhibits some striking features.

\begin{figure*} \centering
    \includegraphics[width=16cm,height=12cm,angle=0]{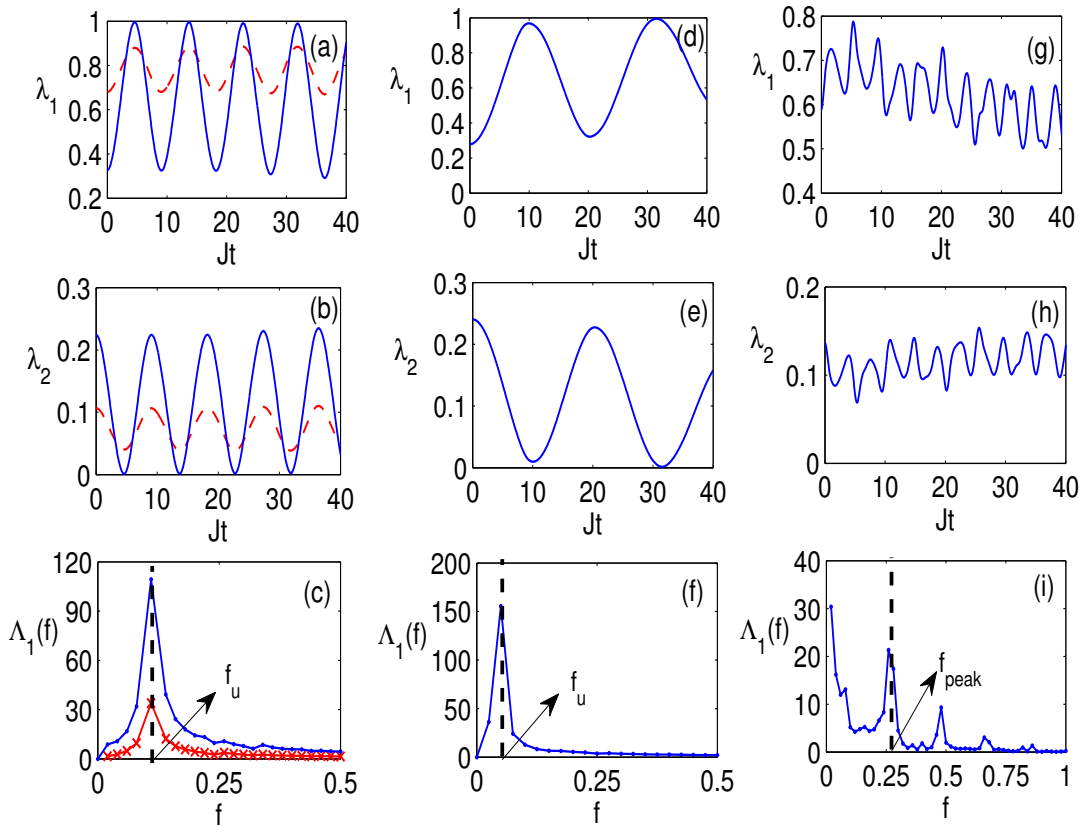}
    \caption{ Quench dynamics of the first two largest Schmidt numbers $\lambda_1$ and $\lambda_2$ versus time for $J'=J'_{opt}=0.3$ in a chain of length $N=20$ with $K_2=0.19$, and $K_1=1$ (blue solid line); $K_1=0.25$ (red dashed line) using exact diagonalization: {\bf plots (a) and (b)}. The Fourier transform $\Lambda_1(f)$ for $J'=J'_{opt}=0.3$ in a chain of length $N=20$ with $K_2=0.19$ and $K_1=1$ (blue line with small solid circles  representing data points) and $K_1=0.25$ (red line with stars representing data points): {\bf plot (c)}.
    Quench dynamics of the first two largest Schmidt numbers $\lambda_1$ and $\lambda_2$ versus time for $J'=J'_{opt}=0.24$ in a chain of length $N=40$ when $K_1=1$ and $K_2=0.08$ using tDMRG: {\bf plots (d) and (e)}. The Fourier transform $\Lambda_1(f)$ for $J'=J'_{opt}=0.24$ in a chain of length $N=40$ with $K_1=1$ and $K_2=0.08$: {\bf plot (f)}. Quench dynamics of the first two largest Schmidt numbers $\lambda_1$ and $\lambda_2$ versus time for a non-optimal case, $J'\neq J'_{opt}$ (here $J'=0.7$), in a chain of length $N=20$ when $K_1=1$ and $K_2=0.19$: {\bf plots (g) and (h)}. The Fourier transform $\Lambda_1(f)$ for $J'=0.7\neq J'_{opt}$ in a chain of length $N=20$ with $K_1=1$ and $K_2=0.19$: {\bf plot (i)}.}
         \label{fig2}
\end{figure*}

\subsection{Entanglement dynamics}
We initially prepare the system in the ground state $|\mbox{GS}(K_1)\ra$ of the Hamiltonian in Eq. (\ref{Ham}), choosing the coupling $K=K_1>K_c$ in the RKKY regime. At time $t=0$ the coupling is instantaneously changed to $K=K_2<K_c$. As a result, the system evolves as
\begin{equation}\label{psi_t}
    |\Psi(t)\ra=\sum_n e^{-iE_n t}\la S_n|\Psi(0)\ra |S_n\ra,
\end{equation}
where $|\Psi(0)\ra = |\mbox{GS}(K_1)\ra$, and where $\{E_n\}$ are eigenvalues of the quench Hamiltonian, defined by Eq. (\ref{Ham}) with $K=K_2$. The corresponding eigenstates $\{|S_n\ra\}$ are global singlets, as implied by spin-rotational symmetry. By tracing part $R$ (see Fig.~\ref{fig1}), one can compute the reduced density matrix $\rho_L(t)$ of part $L$, from which the entanglement spectrum is obtained.

Figs.~\ref{fig2}(a) and (b) show the results for two quantum quenches with different values of $K_1$ using exact diagonalization, with the two largest Schmidt numbers $\lambda_1(t)$ and $\lambda_2(t)$ plotted as functions of time. In both cases $J'\!=\!J'_{opt}$, so that $\xi(J'_{opt},K_2)\!=\!N$. While the amplitudes of the oscillations are different for the two quenches, very surprisingly, the dynamics of $\lambda_{1,2}$ is governed by a single frequency $f$, independent of the quench.  Moreover, as seen in Figs.~\ref{fig2}(a) and (b), the oscillations do not damp.

It is instructive to study the power spectrum
\begin{equation}\label{Lambda_F}
    \Lambda_{n}(f)\equiv \mathcal{F}[\lambda_n(t)] = \frac{1}{\sqrt{2\pi}} \int \lambda_n(t) e^{i2\pi f t} dt, \ n=1,2.
\end{equation}
In Fig.~\ref{fig2}(c), $\Lambda_1(f)$ is plotted for the two given quenches. As revealed by the plot, a sharp peak, independent of the quench, emerges at the frequency $f\!=\!f_u$. This shows that there exists a a unique frequency $f_u$ excited by quenching. In Figs.~\ref{fig2}(d)-(f) the same quantities are plotted for a larger chain, using a time-dependent density-matrix renormalization group (tDMRG) algorithm \cite{white-DMRG-RungeKutta}. Again, single-frequency oscillations are uncovered on the numerically accessible time scale. To a high precision, $\lambda_{1,2}(t)$ can be fit by a sinusoidal function as
\begin{equation}\label{Delta_s_t}
    \lambda_{n}=A_n\sin(2\pi f_u t+\phi_n), \h  n=1,2
\end{equation}
where $f_u=(E_2-E_1)/2\pi$ is the single frequency involved in the dynamics, with $E_{1,2}$ the eigenvalues of the states $|S_{1,2}\ra$ which dominate the expansion in Eq. (\ref{psi_t}). $A_n$ and $\phi_n$ are the quench-dependent amplitude and phase.

\begin{figure*} \centering
    \includegraphics[width=15cm,height=5cm,angle=0]{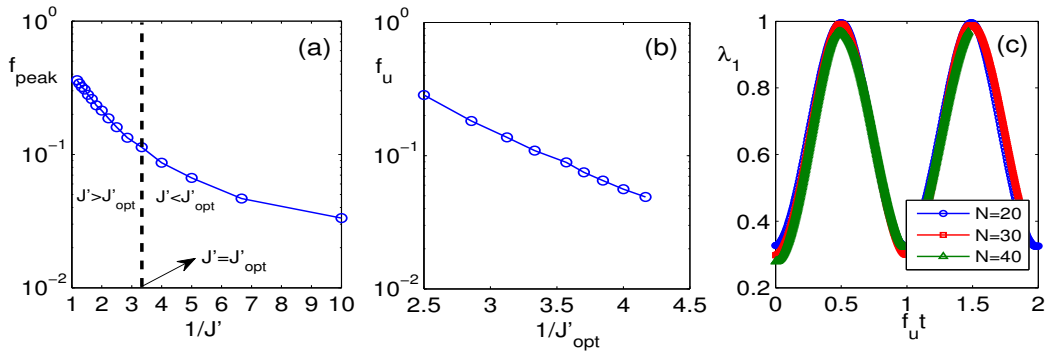}
    \caption{ \textbf{(a)} The dominant  frequency $f_{peak}$ as a function of  $1/J'$ in a semi-logarithmic plot for a chain of fixed length $N=20$ when the impurity coupling $J'$ varies. The deviation from linearity for $J'<J'_{opt}$ is due to the finite size effect as the entanglement length exceeds the system size. \textbf{(b)} The quench-independent frequency $f_u$ as a function of  $1/J'_{opt}$ on a semi-logarithmic plot. Each point corresponds to a different length $N$ for which the optimal coupling $J'_{opt}$ is found and then the quench-independent frequency $f_u$ is determined through time evolution.  \textbf{(c)} Data collapse for the dynamics of $\lambda_1(t)$ as a function of $f_u t$ for three different lengths.}
         \label{fig3}
\end{figure*}

Of course, the single-frequency oscillations disappear for non-optimal Kondo coupling $J' \neq J'_{opt}$ as shown in Figs.~\ref{fig2}(g) and (h). Indeed, these plots clearly display dispersive multi-frequency oscillations, also evidenced by the power spectrum in Fig.~\ref{fig2}(i). It is, however, very important to observe that there always exists a dominant frequency $f_{peak}$ in the power spectrum of $\lambda_{1,2}$ which exhibit a pertinent scaling with  $J'$.
It follows from Fig.~\ref{fig3}(a) that $f_{peak}$ scales with $J'$ as the Kondo temperature,
 \begin{equation} \label{f_peak_Jimp}
f_{peak} \sim T_K ,
\end{equation}
using that $T_K \sim 1/\xi(J',K_2) \sim \exp(-\alpha/J')$ \cite{Jayprakash}.
For $J'=J'_{opt}$, i.e. when $\xi(J'_{opt},K_2)=N$, this frequency becomes the only one accessible to the system, i.e. $f_{peak}=f_u$. For $J'<J'_{opt}$ we see deviations from Eq.~(\ref{f_peak_Jimp}) due to finite-size effects, as  now $\xi(J',K_2)$ exceeds the length $N$.

It is important to realize that the single-frequency oscillations unveiled by our simulations are very different from the quench dynamics studied in Refs. \cite{Eisler-Ent-2007,StephanDubail,Saleur-Kondo-2014,Vasseur-KondoUniversal-2014,Divakaran,Igloi}, which reflects a finite-size effect when instantaneously joining two quantum critical systems. Likewise, the oscillating quench dynamics numerically observed in certain 1D lattice models \cite{Barmettler,Gritsev,Faribault} are also different from our results, coming from a global quench of an integrable interaction, and being either damped \cite{Barmettler} or exhibiting a multi-frequency power spectrum \cite{Gritsev,Faribault}. In all these cases one observes dispersive wave propagation while our scenario is nonperturbative, long-lived and dispersionless.

\subsection{Quench independence}
The frequency $f_u$ is {\em quench-independent}  as it depends neither on $K_1$ nor $K_2$. Independence from $K_1$ is evident from Eq. (\ref{Delta_s_t}). Independence from $K_2$ comes about by tuning the Kondo coupling $J'$ to its ``optimal" value $J'_{opt}$, thus entangling the impurities with {\em all} spins in their respective bulks, i.e. $\xi(J'_{opt},K_2)=N$. For instance, in a chain of length $N\!=\!20$, for $K_2\!=\!0.19$ and $J'_{opt}\!=\!0.30$  one finds that $f_u\!=\!0.11$; the very same value of $f_u$ is obtained for $K_2\!=\!0.1$ and $J'_{opt}\!=\!0.315$.

\begin{figure} \centering
    \includegraphics[width=9cm,height=7.5cm,angle=0]{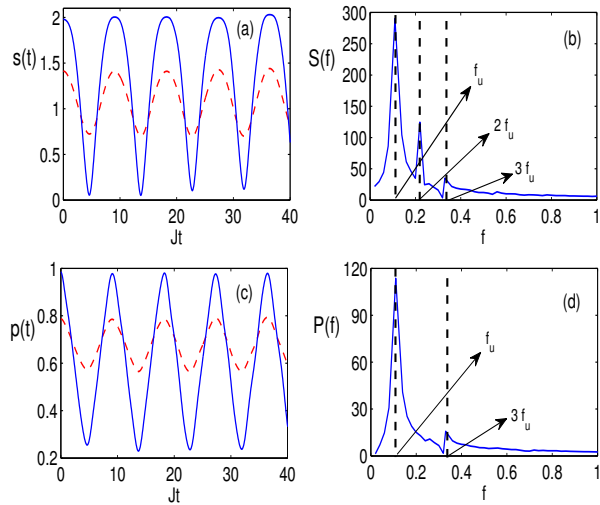}
    \caption{ \textbf{(a)}  The von Neumann entropy $s(t)$ versus time for $J'=J'_{opt}=0.3$ in a chain of length $N=20$ when $K_2=0.19$ and $K_1=1$ (blue solid line) and $K_1=0.25$ (red dashed line). \textbf{(b)} The Fourier transform $S(f)$ of the von Neumann entropy versus $f$ for the case of $K_1=1$. The case for $K_1=0.25$ (not shown in the figure) shows a similar peak at $f_u$ with highly suppressed higher harmonics. \textbf{(c)} The singlet fraction $p(t)$ versus time for $J'=J'_{opt}=0.3$ in a chain of length $N=20$ when $K_2=0.19$ and $K_1=1$ (blue solid line) and $K_1=0.25$ (red dashed line). \textbf{(d)} The Fourier transform $P(f)$ of the singlet fraction versus $f$ for the case of $K_1=1$. }
         \label{fig4}
\end{figure}

The dependence of $f_u$ on the optimal Kondo coupling $J'_{opt}$ is plotted in Fig.~\ref{fig3}(b), showing that
\begin{equation}\label{Fu_Jimp}
    f_u\sim e^{-\alpha/J'_{opt}},
\end{equation}
as expected from Eq. (\ref{f_peak_Jimp}). By combining Eq.~(\ref{Fu_Jimp}) with Eqs. (\ref{Kondo_length_xi}) and (\ref{f_peak_Jimp}), it follows that for a chain of arbitrary size $N$,
\begin{equation}\label{Fu_N}
    f_u\sim \frac{1}{\xi(J'_{opt},K_2)}\, = \frac{1}{N}. 
\end{equation}
To check Eq.~(\ref{Fu_N}), we have computed the product $f_uN$ for various lengths $N$, keeping $K_2$ fixed; see TABLE 1 which shows that $f_uN\simeq 2.1\pm 0.15$. The scaling in Eq. (\ref{Fu_N}) suggests that we get data collapse onto a universal curve for different lengths $N$ if plotting $\lambda_{1,2}$  vs. $f_ut$.  This is confirmed in  Fig.~\ref{fig3}(c).

\subsection{Effective model for $J'=J'_{opt}.$ -}
To expound on the resonance mechanism giving rise to the quench-independent frequency $f_u$, it is crucial to note that, for an optimal quench $J'=J'_{opt}$, only the two singlet eigenstates $|S_{1,2}\ra$ with eigenvalues $E_{1,2}$ are essentially involved in the dynamics, with $\la S_n|\psi(0)\ra\!\simeq\! 0$ in Eq. (\ref{psi_t}) when $n\!>\!2$. Namely, one numerically verifies that there is a small residual overlap
$1- |\la S_1|\mbox{GS}(K_1)\ra|^2 - |\la S_2|\mbox{GS}(K_1)\ra||^2 \!<\! 0.02$ for {\em any} $K_1\! > \!K_c$ and $K_2 \!<\! K_c$ when $J'\! = \!J'_{opt}$. As one moves away from $J'\! =\! J'_{opt}$, other eigenstates rapidly come into play and significantly contribute to the time evolution in Eq. (\ref{psi_t}).

The dominance of two singlet eigenstates $|S_{1,2}\ra$   at the optimal quench $J'\!=\!J'_{opt}$ suggests that the dynamics may be captured by an effective four-spin model. Consider $|\! \!\uparrow \,\ra_{L/R}$ and $|\! \! \!  \downarrow\,\ra_{L/R}$ for the impurity spins in the L/R parts, and $|\! \! \Uparrow \,\ra_{L/R}$ and $|\! \! \Downarrow\,\ra_{L/R}$ for the spins in the L/R bulks (see Fig.~\ref{fig1}). We represent  $|S_1\ra \simeq | 0^- \ra_L \otimes | 0^- \ra_R$, where $|0^\pm\ra = |\! \! \uparrow\Downarrow\, \ra\pm |\! \! \downarrow\Uparrow\, \ra$. Similarly, if we now make the ansatz that $| S_2\ra \simeq | 1 \ra_L \otimes | -\!1 \ra_R - | 0^+ \ra_L \otimes | 0^+ \ra_R + | -\!1 \ra_L \otimes | 1 \ra_R$, with $| 1 \ra = |\! \! \uparrow\Uparrow\, \ra$ and $|-\!1\ra =
|\! \! \downarrow\Downarrow\, \ra$, then the initial state $|\mbox{GS}(K_1)\ra$ will be $| \mbox{GS}(K_1)\ra \simeq \frac{1}{2}|S_1\ra - \frac{\sqrt{3}}{2}|S_2\ra$. Thus, $| \Psi(t)\ra \simeq  \frac{1}{2}|S_1\ra - \mbox{e}^{i 2\pi f_u t} \frac{\sqrt{3}}{2}|S_2\ra$, which periodically brings back $|\mbox{GS}(K_1)\ra$. It is truly remarkable that the quench dynamics of a complex quantum-many body system, when properly tuned, can qualitatively be mimicked by four spins! Intuitively,  the entanglement makes the bulk spins collectively behave as two effective spins, forming a dynamically coordinated composite with the impurities. Putting this intuition on firm ground would be extremely interesting, and could open a new vista on quantum engineered quench dynamics.

\begin{table}
\begin{centering}
\begin{tabular}{|c|c|c|c|c|c|c|c|c|c|}
  \hline
  $N$      & 8     & 12     & 16     & 20    & 24    & 28    & 32    & 36    & 40 \\
  \hline
  $f_uN$ & 2.250 & 2.184 & 2.192 & 2.180 & 2.136 & 2.100 & 2.080 & 2.01  & 2.000 \\
  \hline
\end{tabular}
\caption{ The product $f_uN$ as a function of the length $N$: the table shows that $f_uN\simeq 2.1 \pm 0.15$ independently of the choice of the initial coupling $K_1$ when $K_2 < K_c$.}
\par\end{centering}
\centering{}\label{table_1}
\end{table}

\subsection{von Neumann entropy}
Since effectively one frequency governs the dynamics when  $J'=J'_{opt}$, one may expect that the matrix elements of $\rho_L(t)$ oscillate with a few harmonics of $f_u$.
To verify this, we study the von Neumann entropy  $s(\rho_L(t))$ which depends on all levels of the entanglement spectrum. In  Fig.~\ref{fig4}(a), $s(\rho_L(t))$ is given for two different quantum quenches. When increasing the difference $K_2-K_1$, and thus, the energy released to the system through the quantum quench, deviations from a single frequency sinusoidal function become apparent, with the appearance of higher harmonics in the power spectrum $S(f)=\mathcal{F}[s(t)]$. This is shown in Fig.~\ref{fig4}(b). A similar dependence on higher harmonics of the fundamental frequency $f_u$ is observed for the lower multiplets  of the entanglement spectrum, i.e. for $\lambda_n$ with $n>2$.

\emph{Singlet fraction and spin correlations.-}  The emergence of a single frequency (and of its harmonics) is not a feature only of the dynamics of global quantities such as the entanglement spectrum or the von Neumann entropy. To probe for local quantities we trace out the bulks and compute the reduced density matrix of the two impurities $\rho_{1_L,1_R}(t)$. This has the form of a Werner state due to the spin-rotational symmetry of the model,
\begin{equation}\label{rho_impurity}
    \rho_{1_L,1_R}(t)= p(t) |\psi^-\ra\la \psi^-| + \frac{1-p(t)}{3} \!\sum_{n=0,\pm} |T^n\ra \la T^n|,
\end{equation}
where $p(t)$ is the singlet fraction, $|\psi^-\ra$ is the singlet and $|T^n\ra$ are the triplets. The singlet fraction, which is nowadays experimentally accessible \cite{Petta-QD-2005,Bloch-singlet-triplet-2008}, determines all local properties of the two-impurity composite, such as the two-point correlation $\la \sigma^z_{1_L} \sigma^z_{1_R}\ra=(1-4p)/3$ and the two-impurity concurrence \cite{concurrence} $E(t)={\mbox{max}\{0,2p-1\}}$. In  Fig.~\ref{fig4}(c) the singlet fraction $p(t)$ is plotted versus time for two different quantum quenches. As for the von Neumann entropy, the dynamics can be perfectly matched to the quench-independent frequency $f_u$ for small quenches, and its higher harmonics (essentially the third one) for larger quenches. The Fourier transform $P(f)=\mathcal{F}[p(t)]$ is plotted in Fig.~\ref{fig4}(d), showing the peaks for $f_u$ and $3f_u$.

\section{Conclusions}

In this article we have shown that a local quantum quench across the quantum critical point in a pertinently tuned two-impurity Kondo spin chain may lead to the emergence of long-lived single-frequency oscillations in the dynamics of the entanglement spectrum. The frequency thus revealed is independent from  the energy released by the quench, and also shows scaling behavior with the system size, implying data collapse for the time evolution of levels in the entanglement spectrum.  Important for possible experiments in the future, the quench-independent frequency leaves distinct fingerprints in local observables, and it can be observed as the dominant frequency even when the system is not tuned to make the entanglement length extend over the full system.

The fact that the single-frequency dynamics is found to be tied to the emergence of an optimal entanglement length may hint at new physics. By identifying the entanglement length with the dynamically generated screening length characteristic of Kondo systems \cite{Bayat-TIKM-2012}, our result points towards novel schemes for measuring it in the laboratory. Recent experimental progress in realizing two-impurity Kondo physics with tunable interactions  \cite{bork2011,chorley2012,Spinelli} makes this a tangible challenge.

We conjecture that a single frequency can be associated also to other massless systems. For this to happen there has to be a dynamically generated entanglement length which extends over the full system. In this work we used a spin-chain emulation of the two-impurity Kondo model as a paradigmatic example where this phenomenon occurs. Since the spin-chain emulator is a faithful realization of the spin sector of the two-impurity Kondo model, with a dynamics that effectively decouples from charge at low energies \cite{glazman1999}, we expect that our results will be relevant also for the full two-impurity Kondo model with itinerant electrons. Thus, a spinful double-quantum-dot system, with the confined spins interacting with the conduction electrons in their independent leads, could serve as an experimental setup where the emergence of a long-lived dynamics may be observed. \\



\emph{Acknowledgement:} We would like to thank I. Affleck, F. Buccheri, D. K. Campbell, R. Egger, A. Ferraz, L. Glazman, V. Korepin, K. Le Hur, M. Rigol, and A. Tavanfar for useful discussions. A.B. is supported by the EPSRC grant EP/K004077/1. A.B. also acknowledges the hospitality from the International Institute of Physics (Natal) where this work was initiated.  S.B. acknowledges support of the ERC grant PACOMANEDIA. The visits of H.J. and P.S. to London were supported by ERC grant PACOMANEDIA and EPSRC grant EP/J007137/1 respectively.  H.J. acknowledges support from the Swedish Research Council and STINT. P.S. thanks the Ministry of Science, Technology and Innovation of Brazil, MCTI and UFRN/MEC  for financial support and CNPq for granting a "Bolsa de Produtividade em Pesquisa".  All authors acknowledge the support received from the NORDITA program on "Quantum Engineering of States and Devices" where part of this work was carried out.

\end{document}